\documentclass[aoas,nameyear,dvips]{arximspdf}
\usepackage{graphics}

\doi{10.1214/09-AOAS256}
\volume{3}
\issue{4}
\pubyear{2009}
\firstpage{1566}
\lastpage{1580}

\begin{document}
\begin{frontmatter}

\title{An outlier map for Support Vector Machine classification}
\runtitle{An outlier map for SVM classification}

\begin{aug}
\author[A]{\fnms{Michiel} \snm{Debruyne}\corref{}\ead[label=e1]{michiel.debruyne@ua.ac.be}}
\runauthor{M. Debruyne}
\affiliation{Universiteit Antwerpen}
\address[A]{
Department Wiskunde-Informatica\\
Universiteit Antwerpen\\Middelheimlaan 1G\\ 2020 Antwerp\\ Belgium\\
\printead{e1}} 
\end{aug}

\received{\smonth{4} \syear{2009}}
\revised{\smonth{5} \syear{2009}}

%
\begin{abstract}
Support Vector Machines are a widely used classification technique.
They are computationally efficient and provide excellent predictions
even for high-dimensional data. Moreover, Support Vector Machines are
very flexible due to the incorporation of kernel functions. The latter
allow to model nonlinearity, but also to deal with nonnumerical data
such as protein strings. However, Support Vector Machines can suffer
a lot from unclean data containing, for example, outliers or mislabeled
observations. Although several outlier detection schemes have been
proposed in the literature, the selection of outliers versus
nonoutliers is often rather ad hoc and does not provide much insight in
the data. In robust multivariate statistics outlier maps are quite
popular tools to assess the quality of data under consideration.
They provide a visual representation of the data depicting several
types of outliers. This paper proposes an outlier map designed for
Support Vector Machine classification. The Stahel--Donoho outlyingness
measure from multivariate statistics is extended to an arbitrary kernel
space. A trimmed version of Support Vector Machines is defined trimming
part of the samples with largest outlyingness. Based on this classifier,
an outlier map is constructed visualizing data in any type of
high-dimensional kernel space. The outlier map is illustrated on
4 biological examples showing its use in exploratory data analysis.
\end{abstract}

%
%

\end{frontmatter}

\section{Introduction}

Support Vector Machines [SVM;~\citet{Vapnik08}] are a popular tool for
classification. Two important aspects contributed a lot to this popularity.
First, Support Vector Machines handle high-dimensional, low sample size
data very well, in terms of computational efficiency as well as
prediction quality. Therefore, they are well suited to tackle, for
example, microarray data containing thousands of gene expression levels
(high dimensionality) for a limited number of subjects (low sample
size); see, for example, \citet{Guyon02} and~\citet{Pochet04}.
Second, Support Vector Machines allow for incorporating kernel
functions via the so-called kernel trick. This way nonlinearity in the
data can be handled, for example, using a polynomial or a Gaussian
kernel. Moreover, nonnumerical data can be modeled by designing an
appropriate kernel function using a priori biological information about
the data at hand. This strategy is reported to perform very well, for
instance, in protein homology detection, for example, Fisher SVM~[\citet
{Jaakkola00}], pairwise SVM~[\citet{Liao02}], spectrum kernel~[\citet
{Leslie02}], mismatch kernel~[\citet{Leslie03}] and local alignment
kernel [\citet{Saigo04}].

For high-dimensional and complex data sets, the assumption of
clean,\break
independent and identically distributed samples is not always
appropriate. In \citet{Alon99} and \citet{West01}, for instance, several
samples are~regarded as suspicious. A potential drawback of Support
Vector\break  Machines is the sensitivity to an even very small number of
outliers\break [\citet{Christmann04}; \citet{Steinwart08};\break \citet{Malossini06}].
Outlier detection is thus important and many approaches have been
proposed in the literature. Although often useful, these methods come
with some important drawbacks as well:
\begin{itemize}[$\bullet$]
\item As discussed by \citet{Malossini06}, many techniques are
limited to situations where the sample size exceeds the dimension, thus
excluding modern high-dimensional data analysis.
\item Several types of outliers exist. Algorithms such as those
proposed by \citet{Furey00}, \citet{Li01} and \citet{Malossini06}
focus on samples that are potentially mislabeled. However, not every
outlier is a mislabeled observation and vice versa: a sample can be
correctly labeled yet behave in a completely different way than its
group members. Such discrimination between several types of outliers is
usually not provided.
\item Most algorithms basically provide a ranking of the samples
according to potential mislabeling. However, intuitively it is not
always clear how many of the top ranked samples are serious outlier
candidates. Automatic cut-off procedures often turn out too
conservative (not detecting all outliers) or too aggressive (pointing
out good samples as outliers).
\item The role of the kernel is highly undervalued. Some methods [\citet
{Li01}, \citet{Kadota03}] do not use Support Vector Machines or kernels
at all. \citet{Malossini06} use Support Vector Machines, but
restrict themselves to a linear kernel and even a constant
regularization parameter, whereas optimization of hyperparameters
through cross validation is preferred.
\end{itemize}

In order to avoid some of these difficulties, we propose an outlier map
for~SVM classification. Outlier maps (also called diagnostic plots) are
quite common in multivariate statistics, for example, for linear
regression\break [\citet{Rousseeuw90}] and linear Principal Component
Analysis~[\citet{Hubert04}, \citet{Hubert05}]. The idea is to start from a robust
method guaranteeing resistance to potential outliers. Based on this
robust fit, appropriate measures of interest (e.g., residuals in
regression) are computed and plotted.

In this paper a similar idea is developed for providing an outlier map
which is easy to interpret, distinguishes different types of potential
outliers, and works for any type of kernel. On the $y$-axis of this map
we put the Stahel--Donoho outlyingness. In Section~\ref{sect:sd} we
explain how to compute this outlyingness measure in a general kernel
induced feature space. On the $x$-axis of the outlier map we put the
value of the classification function of a trimmed Support Vector
Machine. More details on this robustified SVM are given in Section~\ref
{sect:sdsvm}. The main part of the paper is Section~\ref{sect:defom},
where the outlier map is defined and illustrated in a simple
two-dimensional example. In Section~\ref{sect:examples} the outlier map is
discussed in $4$ high-dimensional real life examples.

\section{The Stahel--Donoho outlyingness}\label{sect:sd}
Let $Z=\{z_1,\ldots,z_k\}$ be a data set of $d$-dimensional samples
$z_i\in\mathbb{R}^d$.
In multivariate statistics the Stahel--Donoho outlyingness of sample
$z_i$ [\citet{Stahel81}, \citet{Donoho82}] is defined by
%
\begin{eqnarray}\label{eq:stdorig}
r(z_j)=\max_{a\in P}\frac{|a^tz_j-m(a^tZ)|}{s(a^tZ)},
\end{eqnarray}
with $m$ a robust univariate estimator of location and $s$ a univariate
estimator of spread. Popular choices are, for instance, the median for
$m$ and the median absolute deviation (mad) for $s$. The set $P\subset
{R}^d$ is a set of $p$ directions in $\mathbb{R}^d$. In practice, this
set is often constructed by selecting directions orthogonal to
subspaces containing $d$ observations if $d$ is sufficiently small.
Another possibility is taking $p$ times a direction through $2$
randomly chosen observations. This strategy works in any dimension $d$
and since we will extend the outlyingness to high-dimensional kernel
spaces, this is the strategy of our choice. The Stahel--Donoho
outlyingness plays a crucial role in several multivariate robust
algorithms, for example, covariance estimation [\citet{Maronna95}] and
PCA [\citet{Hubert05}].

First we note that this outlyingness measure can be computed in an
arbitrary kernel induced feature space.
Let $\{z_1,\ldots,z_k\}\in\mathcal{Z}$ be $k$ elements in a set
$\mathcal{Z}$. Let $K$ be an appropriate kernel function $K\dvtx \mathcal
{Z}\times\mathcal{Z}\rightarrow\mathbb{R}$ with corresponding feature
space~$\mathcal{H}$ and feature map $\Phi$ such that the inner product
$\langle\cdot,\cdot\rangle$ between feature vectors in~$\mathcal{H}$ can be
computed by $K$:
\begin{eqnarray*}
\langle\Phi(z_i),\Phi(z_j)\rangle=K(z_i,z_j).
\end{eqnarray*}
Denote $\Omega$ the matrix containing $K(z_i,z_j)$ as entry $i,j$. This
matrix is called the kernel matrix. A typical kernel method such as SVM
consists of applying a linear method in the feature space $\mathcal{H}$
such that the computations only depend on pairwise inner products and
thus on the kernel matrix~[\citet{Scholkopf02}]. We now show that the
Stahel--Donoho outlyingness (\ref{eq:stdorig}) can be computed in such
a manner. Let $a$ be the direction in $\mathcal{H}$ through $2$ feature
vectors $\Phi(z_i)$ and $\Phi(z_j)$:
\begin{eqnarray*}
a=\frac{\Phi(z_i)-\Phi(z_j)}{\|\Phi(z_i)-\Phi(z_j)\|}.
\end{eqnarray*}
The projection of a feature vector $\Phi(z_l)$ onto the direction $a$
is then
\begin{eqnarray*}
\langle a,\Phi(z_l)\rangle=\biggl\langle\frac{\Phi(z_i)-\Phi(z_j)}{\|\Phi
(z_i)-\Phi(z_j)\|},\Phi(z_l)\biggr\rangle.
\end{eqnarray*}
Since the squared norm of an element equals the inner product of the
element with itself, we have that
\begin{eqnarray*}
\|\Phi(z_i)-\Phi(z_j)\|&=&\sqrt{\langle\Phi(z_i)-\Phi(z_j),\Phi(z_i)-\Phi
(z_j)\rangle}
\\
&=&
\sqrt{K(z_i,z_i)-2K(z_i,z_j)+K(z_j,z_j)}
\\
&=&\sqrt{(\gamma^{i,j})^t\Omega\gamma^{i,j}}.
\end{eqnarray*}
The vector $\gamma^{i,j}\in\mathbb{R}^k$ denotes the vector with entry
$i$ equal to $1$, entry $j$ equal to $-1$ and all other entries equal
to $0$.
Then
\begin{eqnarray*}
\langle a,\Phi(z_l)\rangle
&=&\biggl\langle\frac{\Phi(z_i)-\Phi(z_j)}{\sqrt
{(\gamma^{i,j})^t\Omega\gamma^{i,j}}},\Phi(z_l)\biggr\rangle
\\
&=&\frac{
K(z_i,z_l)-K(z_j,z_l)}{\sqrt{(\gamma^{i,j})^t\Omega\gamma^{i,j}}}
\\
&=&\biggl(\frac{ \Omega\gamma^{i,j}}{\sqrt{(\gamma^{i,j})^t\Omega\gamma
^{i,j}}}\biggr)_l.
\end{eqnarray*}
Denote $v_{\mathrm{proj}}^{i,j}$ the vector containing the projections of
all feature vectors onto the direction $a$ through feature vectors $\Phi
(z_i)$ and $\Phi(z_j)$:
\begin{eqnarray*}
v_{\mathrm{proj}}^{i,j}=\left(
\matrix{\langle a,\Phi(z_1)\rangle
\cr
\vdots
\cr
\langle a,\Phi
(z_k)\rangle
}
\right)=\frac{ \Omega\gamma^{i,j}}{\sqrt{(\gamma^{i,j})^t\Omega\gamma^{i,j}}}.
\end{eqnarray*}
Note that only the kernel matrix $\Omega$ is needed and not the
explicit feature vectors $\Phi(z_i)$ to compute the projections
$v_{\mathrm{proj}}^{i,j}$.
From these projections the\vspace{1pt} Stahel--Donoho outlyingness of a feature
vector $\Phi(z_j)$ in $\mathcal{H}$ can be calculated as follows:
%
\begin{eqnarray}\label{eq:stdkern}
r(\Phi(z_l))=\max_{(i,j)\in\{1,\ldots,k\}\times\{1,\ldots,k\}}\frac
{(v_{\mathrm{proj}}^{i,j})_{l}-m(v_{\mathrm
{proj}}^{i,j})}{s(v_{\mathrm{proj}}^{i,j})}.
\end{eqnarray}
Again $m$ and $s$ are univariate robust estimators of location and
scale. From this point on we always take
\begin{eqnarray*}
m(v_{\mathrm{proj}}^{i,j})&=&\operatorname{median}(v_{\mathrm{proj}}^{i,j}),
\\
s(v_{\mathrm{proj}}^{i,j})&=&\operatorname{mad}(v_{\mathrm{proj}}^{i,j})=\operatorname
{median}|v_{\mathrm{proj}}^{i,j}-\operatorname{median}(v_{\mathrm
{proj}}^{i,j})|.
\end{eqnarray*}
Note that in~(\ref{eq:stdkern}) we have to check $k(k-1)/2$ directions
to find the maximum, where $k$ denotes the number of observations in
the data set. Then all directions through $2$ observations are
considered. If $k$ is too large, a random subset of directions can be
taken. Typically a few hundred is already enough to provide a good
approximation [\citet{Hubert05}]. In our implementation we use the full
set if $k\leq100$. Otherwise we select $2000$ directions at random.

\section{A simple robust SVM classifier}\label{sect:sdsvm}

\subsection{Algorithm}
Let us now turn to the typical SVM setup. Let $(x_1,\ldots,x_n)$ be a
data set of $n$ training samples in some set $\mathcal{X}$ and let $K$
be a kernel function $\mathcal{X}\times\mathcal{X}\rightarrow\mathbb
{R}$. Let $y_1,\ldots,y_n$ be the corresponding labels: $y_i=-1$ if
sample $i$ belongs to the negative group, $y_i=1$ if sample $i$ belongs
to the positive group. Denote by $n_-$ the number of samples with label
$-1$ and $n_+$ the number of samples with label $+1$.
The following algorithm basically trims a fraction of the data with
largest outlyingness and trains a standard SVM on the remaining
samples. We will refer to this algorithm as SD-SVM (SD stands for
Stahel--Donoho):

\begin{enumerate}[1.]
\item Set $0.5\leq\kappa\leq1$. Denote $h_-=\lfloor\kappa n_- \rfloor
$ and $h_+=\lfloor\kappa n_+ \rfloor$ ($\lfloor c \rfloor$ denotes the
largest integer smaller than $c\in\mathbb{R}$).
\item Trimming step:
Consider only the inputs with group label $-1$.
Compute the Stahel--Donoho outlyingness for every sample in this set
using~(\ref{eq:stdkern}). Retain the $h_-$ observations with smallest
outlyingness. Denote this set of size $h_-$ as $T_-$. Analoguously
obtain the set $T_+$ containing the $h_+$ samples with group label $+1$
with smallest outlyingness.
\item Training step:
 Train a standard SVM on the reduced training set
$T=T_-\cup T_+$. Thus solve
\begin{eqnarray}\label{eq:sdsvmalpha}
&&\max_{\alpha}\sum_{x_i\in T}\alpha_i-\sum_{x_i\in T}\sum_{x_j\in
T}\alpha_i\alpha_j y_i y_j K(x_i,x_j)\nonumber
\\[-8pt]\\[-8pt]
&&\mbox{subject
to}\qquad 0\leq\alpha_i\leq C \quad\mbox{and}\quad  \sum_{x_i\in T}\alpha_i y_i=0.\nonumber
\end{eqnarray}
The classifying function is given by
%
\begin{eqnarray}\label{eq:sdsvmf}
f(x)=\sum_{x_i\in T}\alpha_iK(x_i,x)+b.
\end{eqnarray}
To predict the group membership of a sample $x\in\mathcal{X}$, one
takes $y=\operatorname{sign}(f(x))$.
\end{enumerate}

Note that the computations in the training step are exactly the same as
for an ordinary SVM. The only difference is that the reduced set $T$
containing the observations with smallest outlyingness is used, in
order to avoid negative effects from possible outliers.

\subsection{The regularization parameter}

The regularization parameter $C$ in~(\ref{eq:sdsvmalpha}) is sometimes
set to $C=0.1$ as a default value. However, it is preferable to
optimize the value of $C$. SD-SVM is of course compatible with any type
of model selection strategy: it suffices to add the model
selection
strategy to the training step (step 3) of the algorithm outlined in
Section~\ref{sect:sdsvm}. In all the examples of this paper, 10-fold
cross-validation was used to optimize $C$.

\subsection{Discussion}
\begin{figure}[b]

\includegraphics{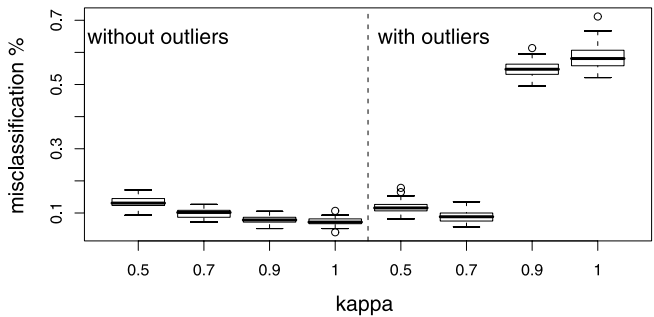}

\caption{Fractions of misclassifications in a small simulation study
for SD-SVM with various values of $\kappa$.}\label{fig:sim}
\end{figure}

To illustrate SD-SVM, consider the following simple experiment: $25$
samples (negative group) are generated, each with $d=1000$ independent
standard normal components. Another $25$ samples (positive group) are
generated with $1000$ independent normal components with mean $0.18$.
In a second setup the same data is used with additional outliers: $4$
samples are added to the negative group with $1000$ independent normal
components with mean $3$. To the positive group, $4$ samples are added
with $1000$ independent normal components with mean $-3$. In both
situations SD-SVM with a linear kernel is applied for several values of
$\kappa\in\{0.5,0.7,0.9,1\}$. The fraction of misclassifications on
$600$ newly generated test data is computed. Figure~\ref{fig:sim} shows
boxplots over $50$ simulation runs.
%
In the case without outliers the number of misclassifications increases
as $\kappa$ decreases. This is quite expected since a lower $\kappa$
means more trimming, which is unnecessary in this case since all
samples are nicely generated from two Gaussian distributions. Thus, it
is no surprise that a classical SVM ($\kappa=1$) performs best.
However, a relatively small amount of outliers (8 out of 58) changes
things completely (right-hand side of Figure~\ref{fig:sim}). A
classical SVM ($\kappa=1$) is no better than guessing anymore (more
than 50$\%$ misclassifications). SD-SVM with $\kappa=0.9$ is not good
enough either, since the trimming percentage is still smaller than the
percentage of outliers. Only if $\kappa$ is chosen small enough, good
performance is obtained. Thus, a small $\kappa$ provides protection
against outliers at the cost of a slightly worse classification
performance at uncontaminated data. For the outlier map it is most
important to avoid the huge effects of outliers, whereas the small
effect of unnecessary trimming is practically invisible. Therefore, a
default choice of $\kappa=0.5$ turns out to be a good choice for the
construction of the outlier map, and we retain this choice throughout
the remainder of the paper.

\section{The outlier map}\label{sect:defom}
\subsection{Construction}
The following visualization is proposed:
\begin{enumerate}[1.]
\item Make a scatterplot of the outlyingness and the value of the
classifier~$f$. Thus, for $j=1,\ldots,n$, plot pairs $(f(x_j),r(\Phi
(x_j)))$ where $r(\Phi(x_j))$ is the Stahel--Donoho outlyingness of
sample $j$ computed in the trimming step of the algorithm and $f(x_j)$
can be calculated from (\ref{eq:sdsvmf}).
\item Plot the inputs with group labels $+1$ as circles and those with
group labels $-1$ as crosses. Add a solid vertical line at horizontal
coordinate~$0$.
\end{enumerate}

\subsection{How to read the map: Toy example}
\begin{figure}[t]
\begin{tabular}{c}

\includegraphics{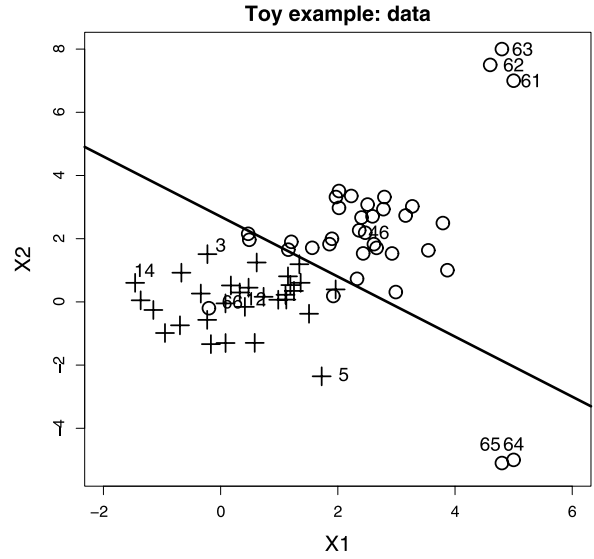}
\\
(a)\\[6pt]

\includegraphics{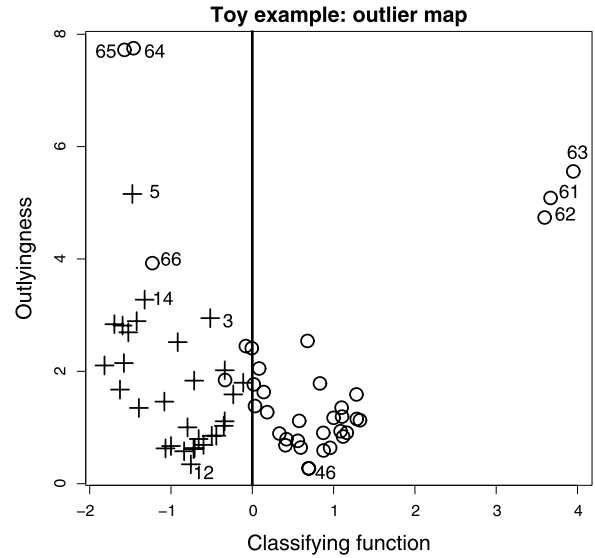}
\\
(b)\\
\end{tabular}
\caption{\textup{(a)} $2$-dimensional classification problem. The solid line
is the SD-SVM classifying line. \textup{(b)}~Corresponding outlier map
visualizing the two main groups and the different types of
outliers.}\label{fig:fig1}
\end{figure}

Consider a simple example in $2$ dimensions as follows: $30$
observations are generated from a bivariate Gaussian distribution with
mean $(0,0)$ and identity covariance matrix. They have group label
$-1$. Thirty observations are generated from a bivariate Gaussian
distribution with mean $(1.5,1.5)$ and identity covariance matrix. They
receive group label $+1$. Apart from these $60$ observations, $6$ more
are added, representing several types of outliers: $3$ data points
(denoted 61--63) are placed around position $(5,7)$ with label $+1$.
Two observations (denoted 64 and 65) with label $+1$ are placed around
$(5,-5)$. One point (denoted $66$) is placed at position $(0,0)$ with
label $+1$. A two-dimensional view of the data is given in Figure~\ref
{fig:fig1}(a).
The solid line represents the SD-SVM classification boundary with a
linear kernel. Despite the $6$ outliers in the data, SD-SVM still
manages to separate both groups quite nicely.

Figure~\ref{fig:fig1}(b) shows the corresponding outlier map. On the
vertical axis one reads the Stahel--Donoho outlyingness. Observations
$12$ and $46$ are positioned in the center of their respective group.
Their outlyingness is indeed small. Observations further away from the
group center have a larger outlyingness, for example, $14$, $3$ and
$5$. On the horizontal axis the value of the classifying function $f$
as in~(\ref{eq:sdsvmf}) can be read. The sign of this function
determines the predicted group labels. The vertical line at $f=0$
divides the plot in two parts: every point left of the line is
classified into the negative group by SD-SVM and every point on the
right is classified into the positive group. We can now see, for
instance, that observation $66$ is a misclassification: it belongs to
the positive group, but receives group label $-1$ since it lies on the
left of the vertical line in Figure~\ref{fig:fig1}(a). The absolute
value of the $x$-coordinate in the diagnostic plot represents a distance
to the classification boundary. In Figure~\ref{fig:fig1}(a) it can be
seen, for example, that observations $3$ and $14$ are almost equally
distant from the negative group center, but observation $3$ is much
closer to the classification line. This information can be found in the
outlier map in Figure~\ref{fig:fig1}(b) as well, since both have
almost the same outlyingness (vertical axis), but $3$ is much closer to
the vertical line than sample $14$ (horizontal axis).

The outliers in the data can be detected and characterized too.
Observations 61--63 are outlying with respect to the other data points
in their group, which is clearly indicated by their large outlyingness.
However, both samples still follow the classification rule. Indeed,
both are lying on the right side in Figure~\ref{fig:fig1}(b). Samples
64 and 65, on the other hand, are outlying with respect to the other
observations in their group as well as with respect to the
classification line: their outlyingness is large and the value of the
classification function is negative, although it should have been
positive to obtain a correct classification. Finally consider
observation $66$. Its not extremely outlying with respect to the other
data points in the positive group. However, taking the negative group
and the classification line into account, it seems to share more
characteristics with the negative group than with its own positive
group colleagues. In the outlier map this is revealed by a moderate
outlyingness and by its position almost in the middle of the left side
of the vertical line.

\section{Examples}\label{sect:examples}
\subsection{Leukemia data}
\begin{figure}[b]

\includegraphics{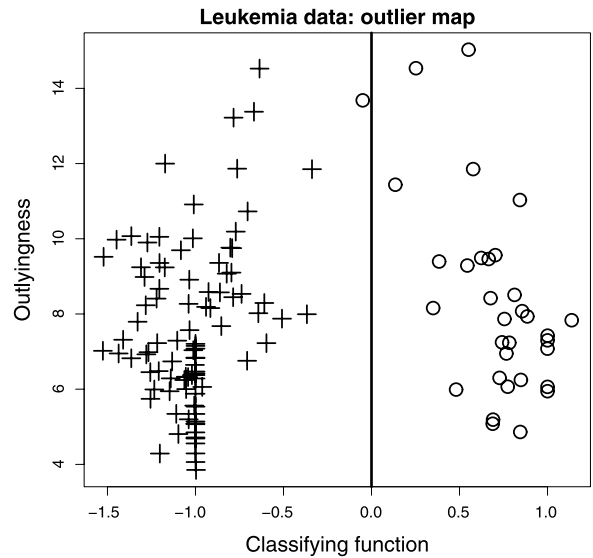}

\caption{Outlier map for the leukemia data. Two nicely separated
homogeneous groups are displayed and one can thus safely proceed
analysis without worrying about outliers.}\label{fig:leuk}
\end{figure}

The first example considers a data set by\break \citet{Chiaretti04}. The data
consist of microarrays from 128 different individuals with acute
lymphoblastic leukemia (ALL), publicly available in the ALL package in
the software environment R. The number of gene expressions at each
individual equals $12625$. There are $33$ adult patients with T-cell
ALL and $95$ with B-cell ALL.
Figure~\ref{fig:leuk} presents the outlier map for SVM with a linear
kernel applied to this data set. It turns out that the data is well
classified and that there are no samples with a very large
outlyingness. Both T-cell and B-cell form homogeneous groups as one
would like when applying a linear SVM. Thus, the outlier map
immediately shows that the data is clean and one can safely proceed
analysis without worrying about outliers.

\subsection{Breast cancer data}
\begin{figure}
\begin{tabular}{c}

\includegraphics{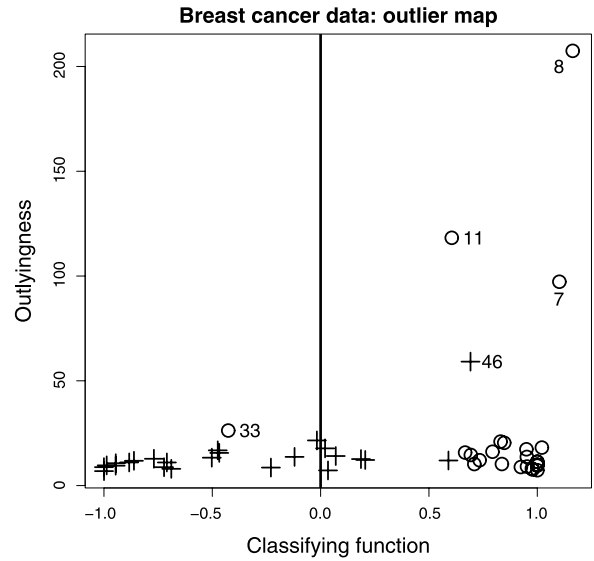}
\\
(a)\\[6pt]

\includegraphics{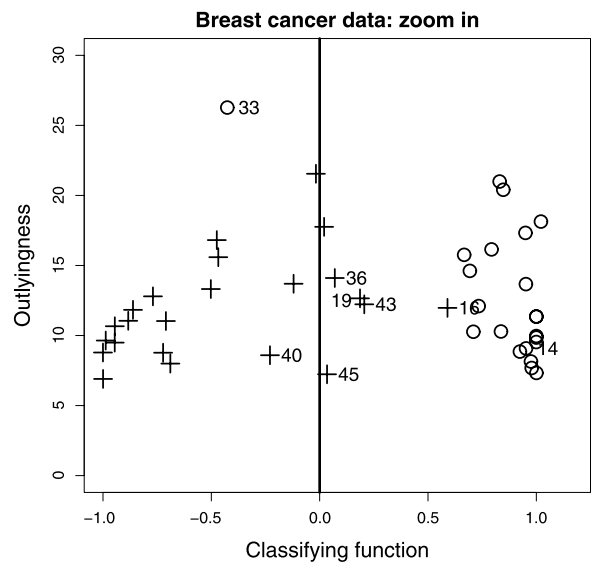}
\\
(b)\\
\end{tabular}
\caption{\textup{(a)} Outlier map for the breast cancer data. Five outliers
are clearly visible. Samples $7,8,11$ are outlying but well classified.
Samples $33$ and $46$ are slightly outlying with respect to their
groups, but are clearly wrongly classified. This suggests that they are
mislabeled rather than erroneous, confirming the original analysis by
West et al. \textup{(b)}~Same plot, but zoomed-in at the region $(0,30)$ on
the vertical axis for better visibility. Observations flagged by
algorithms searching for mislabelings are shown
$(19,36,40,43,45,16)$.}\label{fig:breast}
\end{figure}

The breast cancer data set from \citet{West01} contains $49$ tumor
samples that are either positive (ER$+$) or negative (ER$-$) to estrogen
receptor. The expression levels of $7129$ genes are given for each
sample. For a linear kernel the corresponding outlier map is shown in
Figure~\ref{fig:breast}(a).
Samples $7$, $8$ and $11$ immediately catch the eye. Their outlyingness
is unusually large. In~\citet{West01} samples $7$ and $8$ were already
rejected and taken out of the analysis due to failed array
hybridization. Also sample $11$ was characterized as unusual. It was
the only sample in the ER$+$ group for which the out of sample prediction
was highly unreliable in the analysis performed by~\citet{West01}. The
samples $46$ and $33$ attract attention as well. They have a large
outlyingness and both are clearly misclassified. It turns out that for
this data the group membership ER$+$ or ER$-$ was determined not only by
immunohistochemistry at time of diagnosis, but also by later
immunoblotting. For samples $33$ and $46$ both methods returned
different results. \citet{West01} show via statistical analysis that the
initial labeling ER$+$ for $33$ and ER$-$ for $46$ is probably wrong and
that the immunoblotting results are more appropriate. This is clearly
confirmed by the outlier map.

It is worth noting that the same data set was analyzed in\break \citet
{Malossini06}, where a comparison was made between a proposed stability
criterion, a simple leave-one-out criterion and the algorithm from~\citet
{Furey00}. However, none of these methods was able to detect the $5$
clear outliers discussed so far. Five more suspicious samples were
indicated in~\citet{West01}: $14$, $16$, $40$, $43$ and $45$. In
Figure~\ref{fig:breast}(b) these samples are shown on a zoom-in from
the full outlier map into the region $(0,30)$ on the vertical axis.
Except for $14$, these samples are suspicious in the sense that they
are not confidently classified, since the value of the classifying
function is close to $0$. It is no surprise that these samples are
found by the algorithms compared in~\citet{Malossini06}, since those
methods are designed to detect potentially mislabeled samples. Also
note that some of these mislabeling detection algorithms pointed out
samples $19$ and $36$ as suspicious, although these samples were not
considered in~\citet{West01}. From the outlier map it can be seen that
$19$ and $36$ are indeed wrongly classified by SD-SVM.

\begin{figure}

\includegraphics{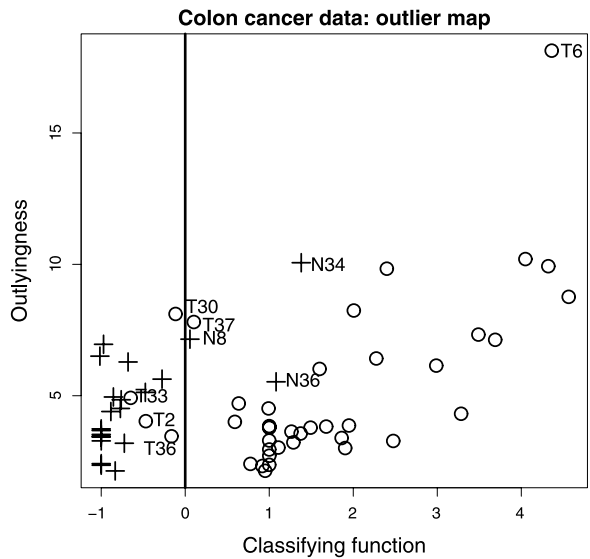}

\caption{Outlier map for the colon cancer data. Misclassifications and
samples with large outlyingness are labeled. These were also flagged in
the original analysis by Alon et al. (\protect\citeyear{Alon99}).}\label{fig:alon}
\end{figure}
%
\begin{figure}[t]

\includegraphics{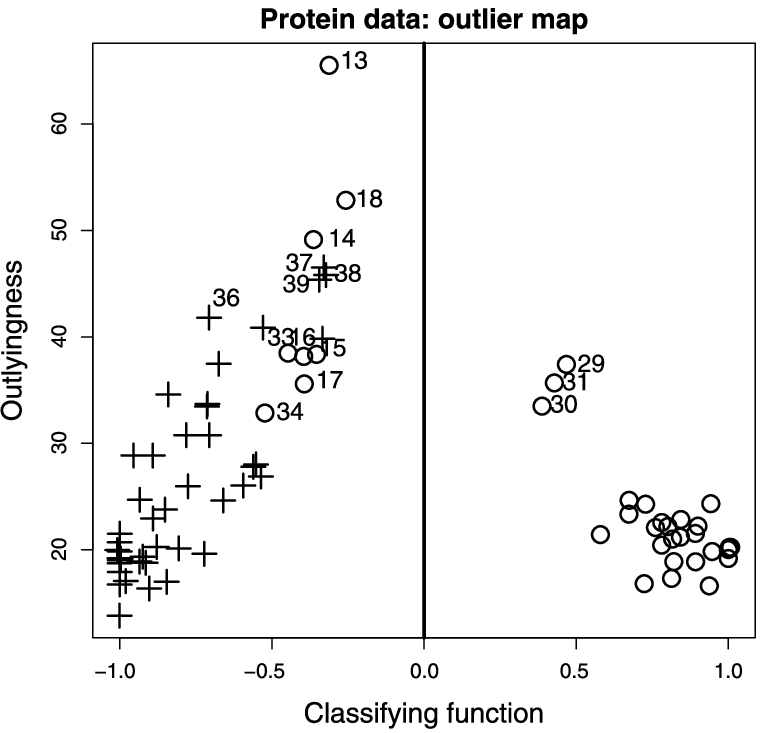}

\caption{Outlier map for the protein data. The heterogeneity of the
positive group is clearly visible, with different clusters according to
the subgroups of different phyla, also confirming the original
clustering analysis by Pollack et al. (\protect\citeyear{Pollack05}).}\label{fig:protein}
\end{figure}

\subsection{Colon cancer data}
The colon cancer data set from~\citet{Alon99} contains $2000$ gene
expression levels for $40$ tumor samples and
$22$ normal samples.
The outlier map with a linear kernel is shown in Figure~\ref{fig:alon}.
In the tumor group T2, T33, T36 and T30 are misclassified. Sample T37
is classified correctly, but with low confidence: it is very close to
the classification boundary. In the normal group N8 and especially N34
and N36 are the suspicious cases that behave differently from the other
normal samples. The 8 aforementioned samples plus sample N12 were
identified as possible outliers in the original paper by~\citet{Alon99}
for biological reasons. Thus, 8 out of 9 true outliers can be
identified on the outlier map, only leaving N12 undetected. However,
in~\citet{Malossini06} none of the methods that were compared could
detect N12. Moreover, the stability criterion proposed by Malossini et
al. was unable to detect T37 and N8 too and incorrectly pointed at N2
and N28 as possibly suspicious samples.
Also note the interesting sample T6. From the outlier map we see that
this sample is classified correctly and with much confidence.
Nevertheless, its outlyingness with respect to the other tumor samples
is rather large. This means that T6 behaves quite differently than the
other tumor samples, but without distorting the classification. In
Malossini et al. most of the methods analyzed did not detect T6 at all.
Again this is no surprise since methods such as the stability criterion
of Malossini et al. specifically focus on mislabeled observations,
whereas T6 is certainly not mislabeled. Only the outlier detection
method of~\citet{Kadota03} is able to detect T6, but does rather poorly
on the other samples detecting only 5 out of 9 true outliers.

\subsection{Protein data}
The protein data set taken from~\citet{Pollack05} contains $131$ protein
sequences of the essentially ubiquitous glycolytic enzyme
3-phosphoglycerate kinase (3-PGK) in three domains: Archaea, Bacteria
and Eukaryota.
The data set is available in the Protein Classification Benchmark
Collection at \url{http://net.icgeb.org} (accession number PCB00015). We
consider here classification task number 10 where the positive group
consists of 35 Eukaryota. The negative group consists of 4 Archaea and
40 Bacteria. To classify these two groups of protein sequences, we use
SVM with the local alignment kernel~[\citet{Saigo04}]. Default parameter
values were used: gap opening penalty $=$ 11, gap extension penalty $=$ 1,
scaling parameter $=$ 0.5. The outlier map is shown in Figure~\ref
{fig:protein}. One observes that the positive group of Eukaryota is
very heterogeneous as several clusters appear. These clusters all have
a biological interpretation, as the group of Eukaryota contains several
subgroups of different phyla. For instance, observations 29--31 are from
the phylum of Alveolata. Samples 13--17 are the Euglenozoa. Note that 18
(named Q8SRZ8), which belongs to the Fungi, was clustered in the group
of Euglenozoa by Pollack et al.; this is actually confirmed by the
outlier map. Finally, samples 33 and 34 are outlying with respect to
the positive group. They form, together with 32, the group of
Stramenopiles. Note that the different behavior of sample 32 from its
fellow Stramenopiles is again a confirmation of the analysis by Pollack
et al.: their clustering method assigned 32 (named Q8H721) in the main
group of Eukaryota Metazoa. Also, in the outlier map 32 is situated in
the main group, whereas 33 and 34 form a separate cluster.
In the positive group the heterogeneity is less clear, although the 4
Archaea (36--39) do have the largest outlyingness compared to the other
samples which are all Bacteria.

\section{Conclusion}

An outlier map is proposed for Support Vector Machine classification.
If the outlier map shows two homogeneous and well classified groups,
one can safely proceed analysis without worrying about outliers.
However, in some situations this may not be the case and the outlier
map can be a simple and useful tool to detect this. Moreover, the
outlier map can be drawn for any choice of kernel, including rather
exotic ones such as used in protein analysis. It can also be helpful to
gain insight in the type of outliers, for example, whether outliers are
mislabeled observations or not, or whether the outliers are isolated
errors or rather a small subgroup of the group structure considered.
This is important to know how to proceed analysis. If the outliers are
truly erroneous observations, one should not take them into account to
build a classifier, and one can manually discard them from the data set
or apply a robust classifier. If the outliers are mislabeled
observations, one probably should re-examine the labeling and change
the label of the outlier if this seems indeed appropriate. If the
outliers form a small subgroup of the data, one might reconsider the
use of a binary classifier and turn to a more appropriate modeling technique.
In any event, the outlier map can be helpful for practitioners of SVM
classification to make such decisions.

\printaddresses

\end{document}